\documentclass[a4paper]{article}

\usepackage{INTERSPEECH2021}
\usepackage{tipa}
\usepackage{multirow}

\title{Synchronising speech segments with musical beats in Mandarin and English singing}
\name{Cong Zhang$^1$, Jian Zhu$^2$}
\address{
  $^1$Radboud University\\
  $^2$University of Michigan}
\email{cong.zhang@ru.nl, lingjzhu@umich.edu}

\begin{document}

\maketitle
\begin{abstract}
Generating synthesised singing voice with models trained on speech data has many advantages due to the models' flexibility and controllability. However, since the information about the temporal relationship between segments and beats are lacking in speech training data, the synthesised singing may sound off-beat at times. Therefore, the availability of the information on the temporal relationship between speech segments and music beats is crucial. The current study investigated the segment-beat synchronisation in singing data, with hypotheses formed based on the linguistics theories of P-centre and sonority hierarchy. A Mandarin corpus and an English corpus of professional singing data were manually annotated and analysed. The results showed that the presence of musical beats was more dependent on segment duration than sonority. However, the sonority hierarchy and the P-centre theory were highly related to the location of beats. Mandarin and English demonstrated cross-linguistic variations despite exhibiting common patterns.
\end{abstract}

%tipa reference: https://www.tug.org/TUGboat/tb17-2/tb51rei.pdf

\section{Introduction}
%\subsection{Singing Synthesis}

The world-wide success of virtual singers, such as Hatsune Miku, has generated widespread interest in further development and research into synthesised singing. Over the course of the past few decades, the advancement of text-to-speech algorithms has immensely improved the performance of synthesised singing. The state-of-the-art models make use of deep learning methods (e.g. \cite{Gu2020,Ren2020}), which generate much more natural singing voice than the traditional concatenative or parametric models. However, deep learning models usually require more resources to build and to run. These shortcomings suggest that the lightweight and flexible traditional parametric methods are still useful in specific settings. Among all alternatives, optimising existing text-to-speech models trained with speech data for the use of singing voice synthesis seems to be still relevant. 

Admittedly, compared with deep learning models, speech-based models generate singing voices that are less natural; they nevertheless allow a wider community to access the resources and create their own singing voice. The majority of the issues exist due to the lack of alignment between musical information and speech information. Among all issues induced by the missing information, the temporal alignment between speech segments and musical beats strikes as the most serious issue: the synthesised singing sounds off-beat at random and thus sounds unnatural to human ears. This is because musical beat information is not present in the speech training data, and therefore cannot be acquired through machine learning. This information thus needs to be provided to the models in other ways. Previous studies have adopted various methods in resolving this issue. \cite{Blaauw2017,valle2020mellotron} used a heuristic way of assigning duration values or fittings to different segments (e.g. setting one phoneme to 20 ms, and another to 100 ms). \cite{Saino2006} proposed an HMM-based ``time-lag" model that was trained on 60 Japanese songs to account for the temporal difference between the musical notes and phonemes. With the addition of the time-lag model, the mean opinion score (MOS) was substantially improved. In more recent studies, the importance of duration was also highlighted and modelled using various deep learning models (e.g. \cite{valle2020mellotron, Zhang2020,Hono2019}). However, these duration models still required a substantial amount of data for model training. This study aims to deliver an interdisciplinary analysis using linguistic and music knowledge, to inform future singing voice synthesis model building. The results can, for instance, replace the heuristic values used in such studies as \cite{Blaauw2017,valle2020mellotron}. By providing a way of enhancing the temporal features, this study can also be beneficial to the more complicated modelling methods. It can also supplement the current modelling algorithms with domain knowledge, without attempting to fit any predictive synthesis models. An additional goal of this study is to gain insights into speech rhythm through analysing singing data that contain speech segments and fixed rhythm.

\subsection{Linguistics Background}
\textbf{Speech Rhythm.}
While speech is not guided by musical scores, it encodes a natural rhythm. Many past studies have examined how speech rhythm was produced and perceived. One of the most relevant theories is the ``P-centre" theory, i.e. perceptual centre. It was proposed to account for the alignment of timing units and speech segments \cite{morton1976perceptual}. Both acoustic and perceptual studies (e.g. \cite{marcus1981acoustic, hoequist1983perceptual, cooper1988syllable, chow2015syllable}) have found the factors influencing the locations of P-centres to be the phonological structure of syllables. In a syllable consisting of a syllable-initial consonant and a subsequent vowel, the P-centre is usually close to the vowel onset  \cite{de1992acoustic,patel1999acoustics,barbosa2005abstractness}. \cite{barbosa2005abstractness} has also found the energy profile of syllable-initial consonants to be important to the P-centre location. Many other traditional speech rhythm metrics such as $\%V$, $\Delta V$ and so forth are also tied to the timing of vowels. Recent studies employed more up-to-date metrics such as \texttt{maxE} and \texttt{maxD}, and also concluded the importance of vowel onsets and energy in the synchronisation of speech rhythm \cite{Rathcke2021}. \\[0.15cm]
\noindent\textbf{Sonority Hierarchy.}
Another relevant topic is the sonority hierarchy of speech segments \cite{Pike1943}. Many studies have looked into the sonority of different segment types, and the most widely agreed version of the hierarchy is: \texttt{vowels $>$ glides $>$ liquids $>$ nasals $>$ obstruents} \cite{Clements1990}. While the more sonorant types (vowels, glides, liquids, nasals) attracted more agreement, the order of the three obstruent types (stops, fricatives, and affricates) has been much more controversial. \cite{parker2002quantifying} studied the acoustic correlates and found that while intensity contributed to the majority part of the sonority hierarchy, other parameters such as formant, duration, and peak air flow all served as relevant factors. Sonority hierarchy is largely universal; however, minor cross-linguistic variations also exist.

\subsection{The Current Study}
We analysed a Mandarin singing corpus and an English singing corpus. The two languages were chosen because, according to tradition rhythmic class theories, Mandarin and English were considered as syllable-timed and stressed-timed respectively \cite{Mok2009, Grabe2008}. While we have reservations about the theoretical validity of this classification, it offers a convenient division to account for the typological differences between the two languages in terms of syllable structures, vowel reduction rules, etc. The following research questions and hypotheses were addressed: 

{\it 1. What types of speech units attract most of the beats?} Since we expect the alignment to favour vowels, it is likely that other sonorant phoneme types also anchor musical beats. We hypothesise that the more sonorant a segment is, the more likely it is to have beats aligned with it. 

{\it 2. Where do musical beats align with speech segments?} The theories related to speech rhythm such as P-centre theory suggest that the vowel onsets and the transition between the preceding consonants and the vowels are more likely to be the synchronising points for speech rhythm. We therefore hypothesise that the rimes in Mandarin and the vowels in English are aligned with the most beats, and the beat locations are likely to be at the beginning of the vowels or the end of preceding consonants. Moreover, the more sonorant the segments are, the more vowel-like they are, and thus the earlier the alignments are. 

{\it 3. Are there cross-linguistic differences?} We expect Mandarin and English to exhibit different synchronisation patterns. Given that sonority hierarchies can exhibit language-specific variations, and English and Mandarin belong to different linguistic typologies in many respects, the two languages are likely to exhibit speech-beat synchronisation differences.

\section{Mandarin Singing Analysis}
\subsection{Data}
\textbf{Corpus.} This study used a corpus of 33 excerpts of singing from 23 songs, with each excerpt having a different tempo. All songs were Mandarin pop songs that were selected to have similar styles. A professional female singer's unaccompanied singing was recorded in a professional recording studio, with a professional sound engineer monitoring the quality of the singing. The recordings were made at a sampling rate of 48 kHz at 32-bit in mono channel. The total length of the songs in this corpus was 111.5 minutes.\\[0.15cm]
\noindent\textbf{Data Annotation}
The musical beats and the linguistic information were manually annotated by two different annotators. For beat annotation, a music expert with choral conducting experience annotated the beats by adding metronome beats in GarageBand \cite{garageband} for each excerpt. They noted down the duration from the onset of the audio to the first beat, as well as its tempo (i.e. beats per minute). Two other annotators with experience of playing music instruments listened to the audio files with metronome beats. Disputed parts were returned to the music expert to re-annotate or excluded when no consensus could be achieved. For the linguistic segmentation, a student assistant manually segmented all Mandarin syllables into onsets and rimes in all 23 songs. The assistant received daily training in phonetic segmentation. They were trained to follow the principles in \cite{Machac2009}. An experienced phonetician checked all the practice annotations during the training, and also cross-checked 20\% percent of the final corpus annotation to ensure the accuracy and consistency of the segmentation. The musical and linguistic segmentation processes were kept independent from each other to avoid any subjective presumptions from the annotators. The final annotations were integrated into Praat \cite{Boersma2020} TextGrids. % as shown in Figure~\ref{fig:annotation}.
The onsets and rimes that aligned with musical beats were labelled as \texttt{beat}, and the ones that did not were labelled as \texttt{no beat}. The location of a beat was calculated as a percentage indicating the relative position of the beat within a segment with respect to the total segmental duration. Onsets and rimes are important units in Mandarin and often function as a whole structure (e.g. Tone-Bearing Unit) in organising phonology, speech planning, and speech perception. The onsets and rimes in this study were categorised by the phoneme types of the onsets and the syllable structure of the rimes. The phoneme types included fricative, affricate, nasal, stop, liquid, and glide. The syllable structure for the rimes in Mandarin included V (monophthong), VV (diphthong), VN (Vowel + nasal coda), G (glide) can also appear before these structures and were also counted as a part of the rimes (for more information about Mandarin syllable structure and phoneme types, see \cite{Duanmu2000e} \cite{Zhangcong2020}). 

%in relation to the starting time of the onsets and rimes: 
%\texttt{Beat Location = (time of beat - starting time of the segment) / duration of the segment}. 

\subsection{Results and Discussions}
\begin{figure}
    \centering
    \includegraphics[width=0.48\linewidth]{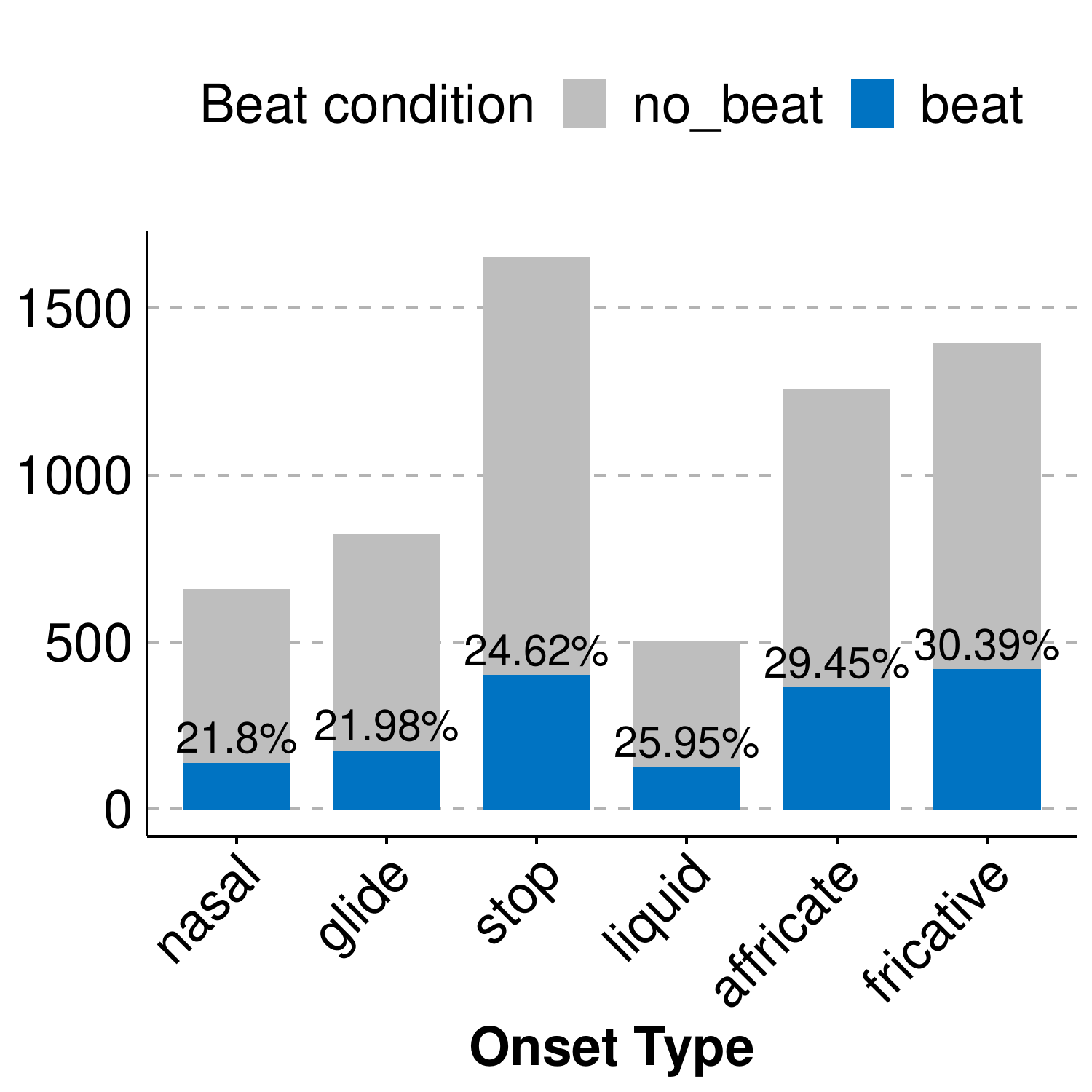} 
    \includegraphics[width=0.48\linewidth]{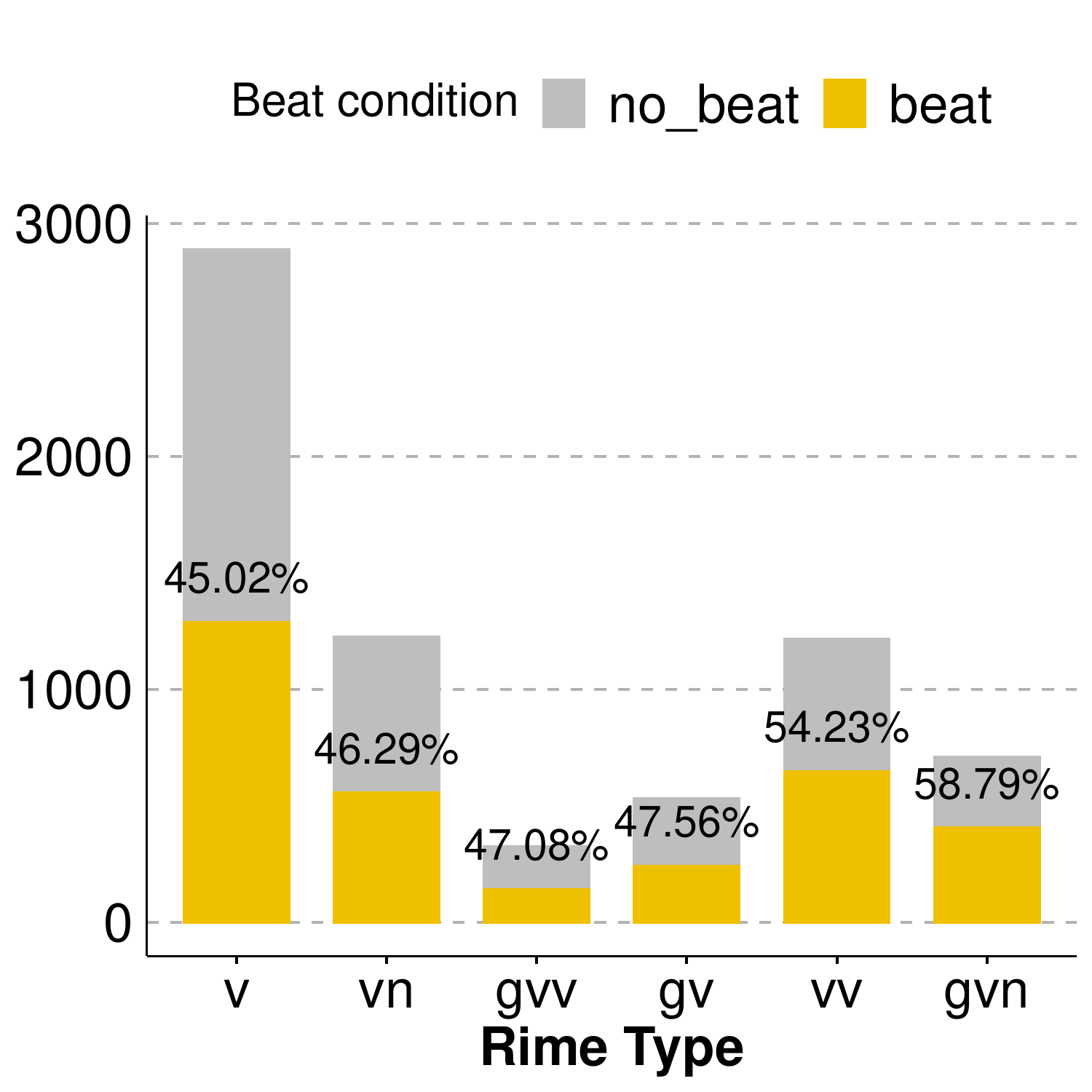} 
    \caption{The numbers and proportions of beat-aligned onsets ({\bf left}) and rimes ({\bf right}) by onset and rime types. Y axis: number of onsets or rimes. %Blue/yellow bar: proportion of segment types that have beats aligned; grey bar: proportion of segment types that have no beat aligned. The bars are ranked by the percentages from low to high. 
    }
    \label{fig:cn_bar}
\end{figure}

\begin{figure}
    \centering
    \includegraphics[width=0.48\linewidth]{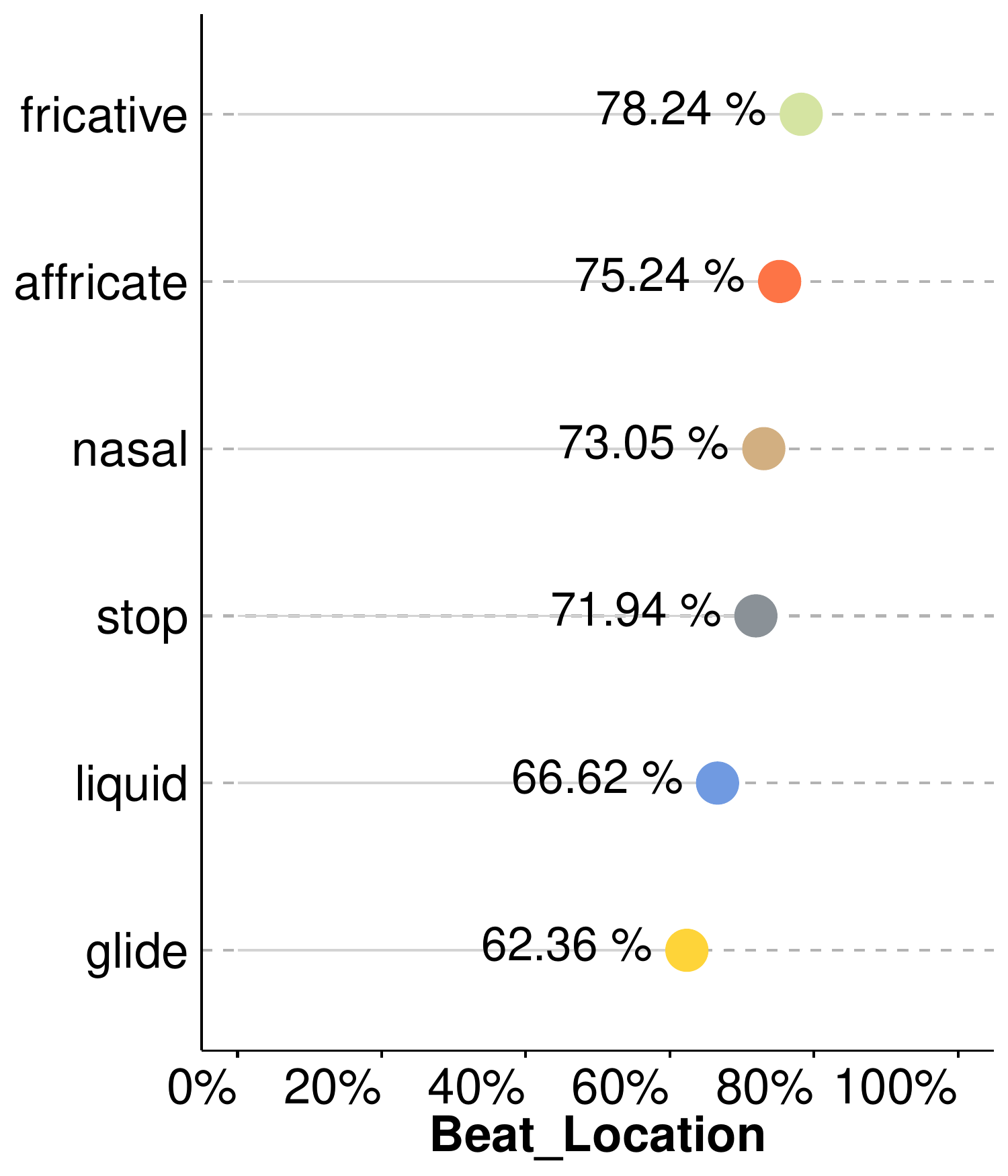} 
    \includegraphics[width=0.48\linewidth]{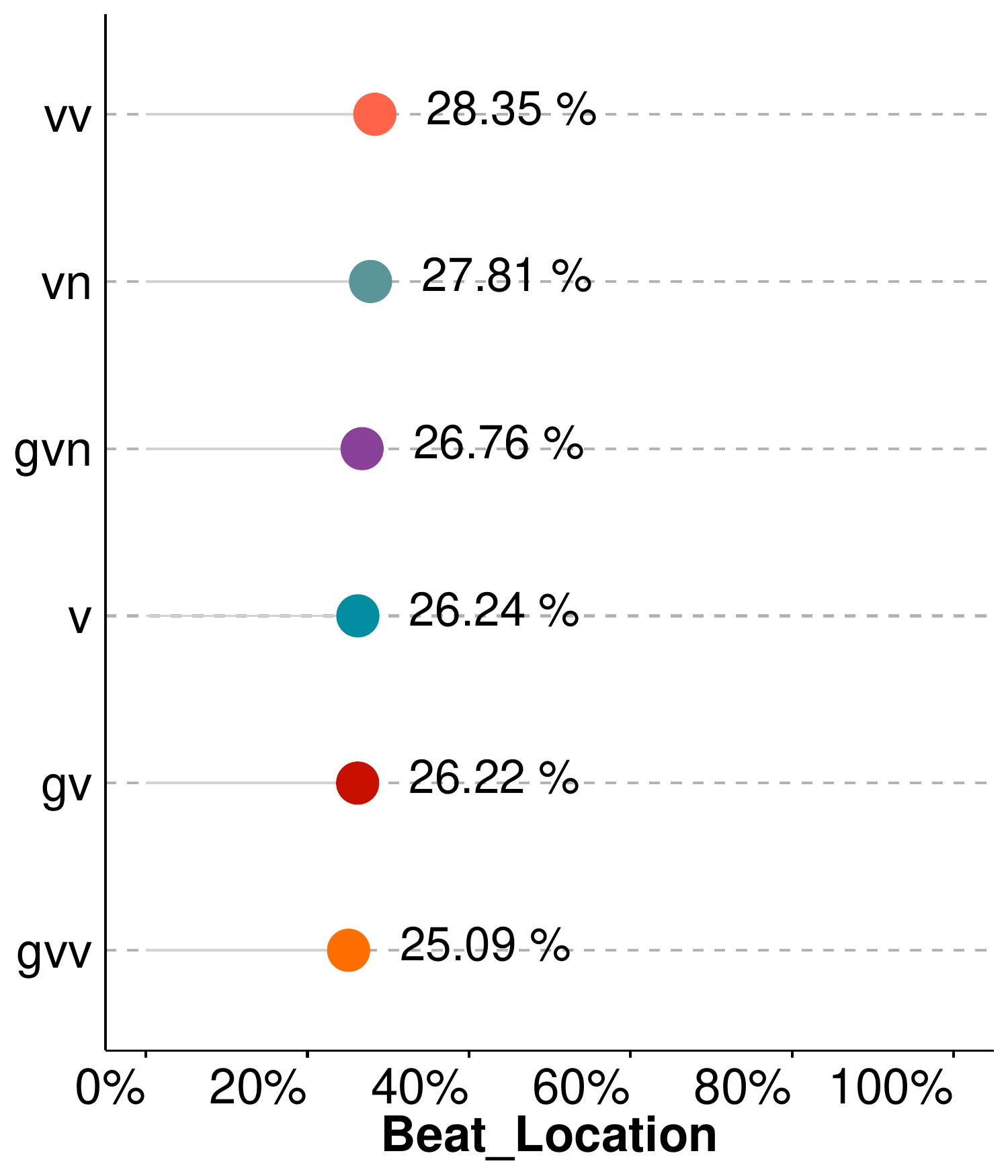} 
    \caption{The average location of beat in relation to the aligned onsets ({\bf left}) and rimes ({\bf right}).}
    \label{fig:cn_lolli}
\end{figure}

\textbf{Presence of Beats.}
The Mandarin dataset consisted of 6653 Chinese characters, among which 65\% were aligned with musical beats. The total number of onset and rimes were 6270 and 6902 respectively. Only 26.3\% of onsets (1651) received beats, while almost half of the rimes (3353, 48.6\%) aligned with beats. In Figure~\ref{fig:cn_bar}, the types of onsets and rhymes were ranked from low to high by the proportion of beat-aligned onsets and rimes. Among all the onset types, beat-aligned nasals accounted for the smallest proportion, while the fricatives had the largest proportion. The results exhibited a pattern contrary to the sonority hierarchy. Nasals, glides, and liquids are high on the sonority scale, but a smaller proportion of these phoneme types had beat-alignment. The non-sonorant affricates and fricatives, however, had higher chances of aligning with beats. This is against our hypothesis 1 which was based on the sonority hierarchy. However, a possible explanation was that sonority hierarchy was overridden by a more important factor. In this case, phoneme duration was likely to be the more important factor: affricates and fricatives are generally much longer than the other types of the consonants; therefore, there were higher chances for beats to fall onto them. We will examine the effect of phoneme duration on predicting beat presence in Section 4.\\[0.15cm]
\noindent\textbf{Location of Beats.}
Figure~\ref{fig:cn_lolli} displays the beat location for different phoneme types of the onsets (left) and by different syllable structures for the rimes (right). The onsets showed a strong correlation with the sonority hierarchy in that the sonorant onsets such as glides and liquids had beats aligned to almost the mid point of the onsets. However, the alignment became very late towards the end for the non-sonorant onsets such as fricatives and affricates. For the vowels, alignments across all types were at around the first 25\% of the rimes. These results support our hypothesis 2 because the beats did align with the vowel onsets and the transition between the preceding consonants and the vowels; moreover, the data illustrated that the more sonorant the segments were, the more vowel-like they were, and thus the earlier the alignments were.

\section{English Singing Analysis}
\subsection{Data}
\textbf{Corpus.} The English singing corpus contained 18 excerpts of singing from 14 English songs, each having a different tempo. The nature of this study required data from professional singers to ensure the beats were correctly produced. However, unaccompanied professional singing data were extremely difficult to acquire. Therefore, we used as many songs as we could access. The data were saved from YouTube as \texttt{.wav} files. While the compressed YouTube files may not be suitable for rigorous acoustic analysis, this study only focuses on the timing of the beats in relation to the segments, we therefore considered YouTube audios sufficient for the purpose. The total length of the audio files was 103.4 minutes. \\[0.12cm]
\noindent\textbf{Data Annotation.}
The same annotators annotated the English singing data following the same procedures. The only difference was the unit of linguistic annotation: in English, consonants and vowels were used instead of onsets and rimes, since onsets and rimes were rarely important units in English. The consonants were categorised into the seven phoneme types according to the characteristics of English: approximant (including glides and liquids), nasal, voiced and voiceless stops, voiced and voiceless fricatives, and affricate. Vowels were divided into monophthongs and diphthongs, which corresponded to V and VV categories in the Mandarin dataset.

\subsection{Results and Discussions}
\begin{figure}
    \centering
    \includegraphics[width=0.48\linewidth]{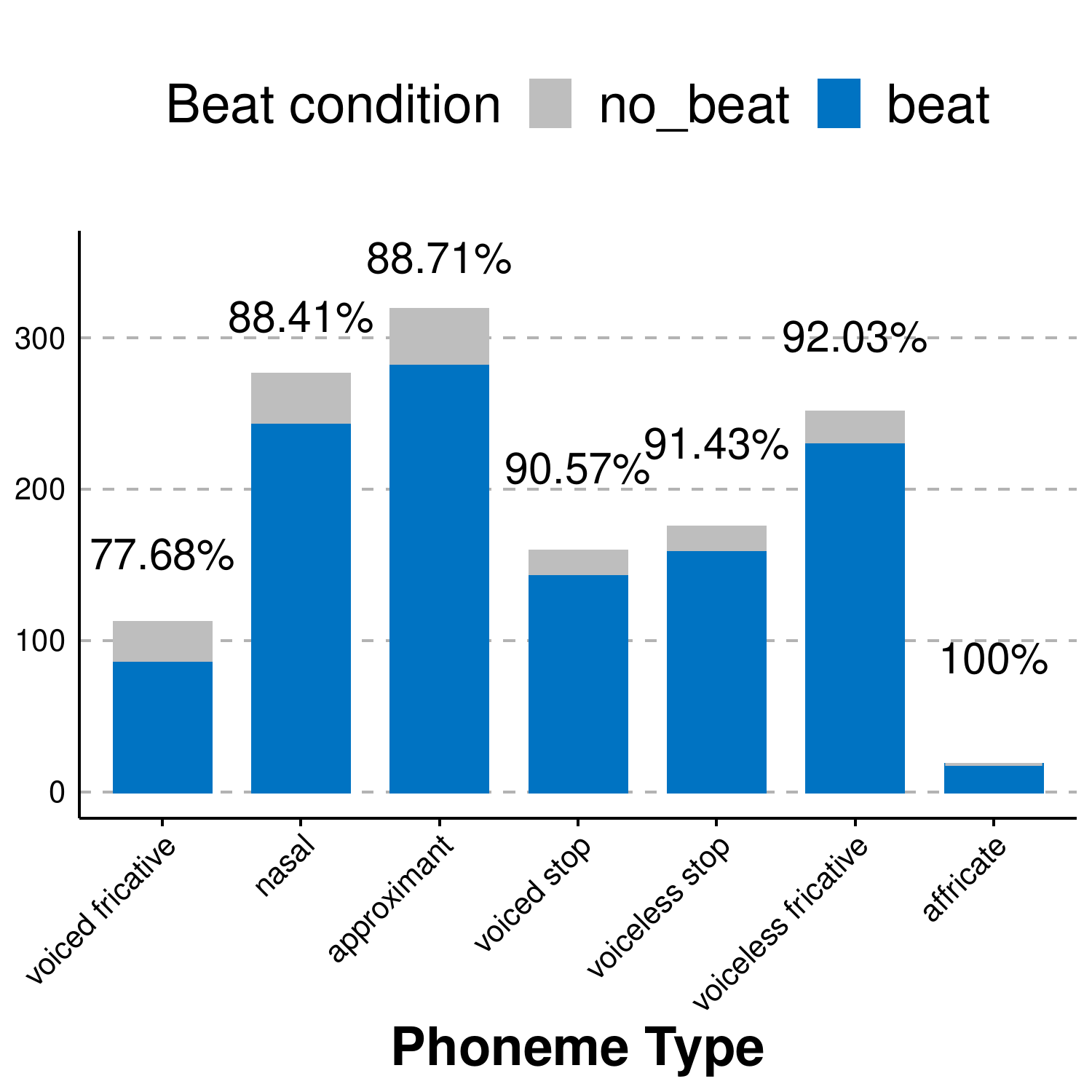}
    \includegraphics[width=0.48\linewidth]{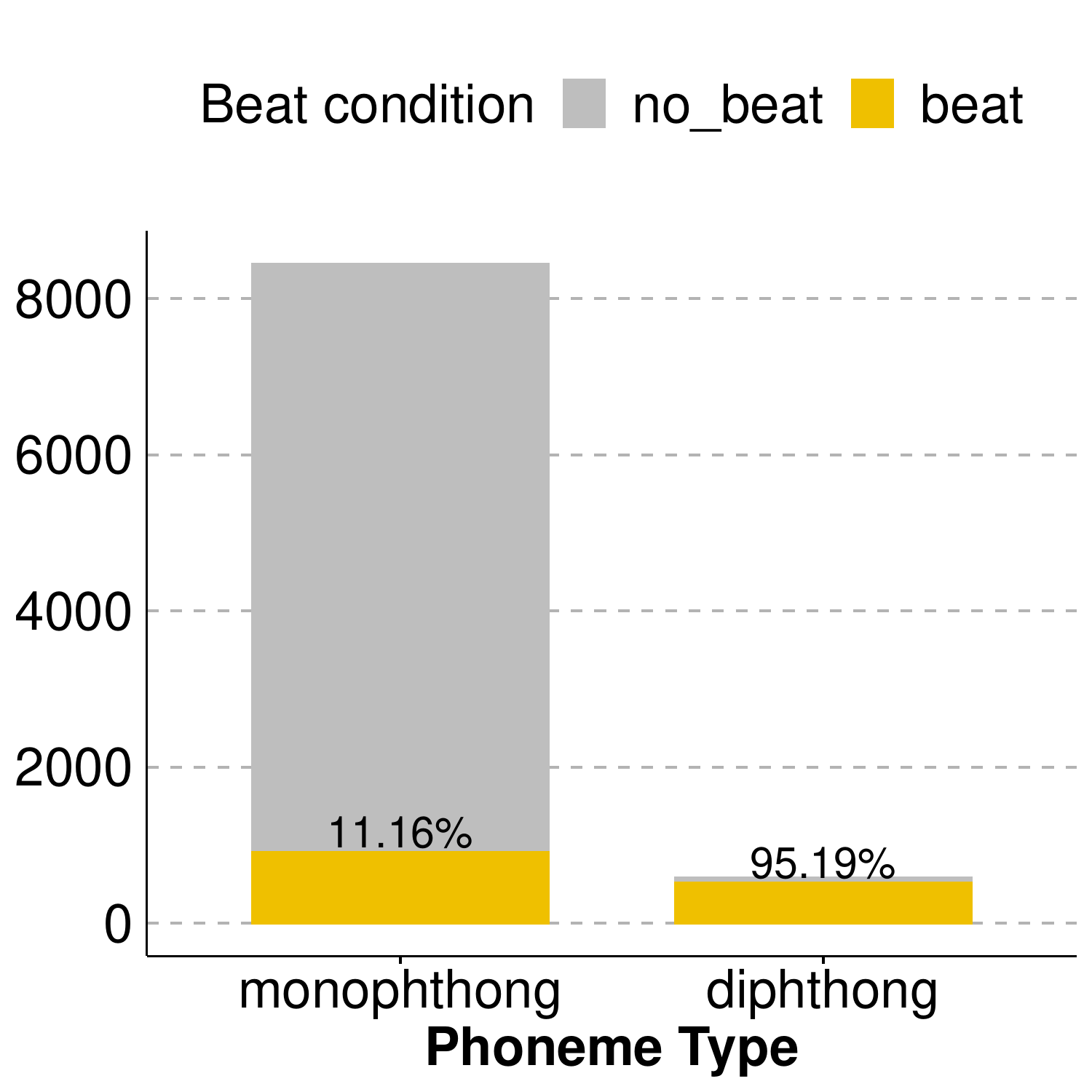}
    \caption{The numbers and proportions of beat-aligned consonants ({\bf left}) and vowels ({\bf right}) by phoneme types. Y axis: number of consonants or vowels. %Blue/yellow bar: proportion of segment types that have beats aligned; grey bar: proportion of segment types that have no beat aligned. The bars are ranked by the percentages from low to high.
    }
    \label{fig:en_bar}
\end{figure}

\begin{figure}
    \centering
    \includegraphics[width=0.48\linewidth]{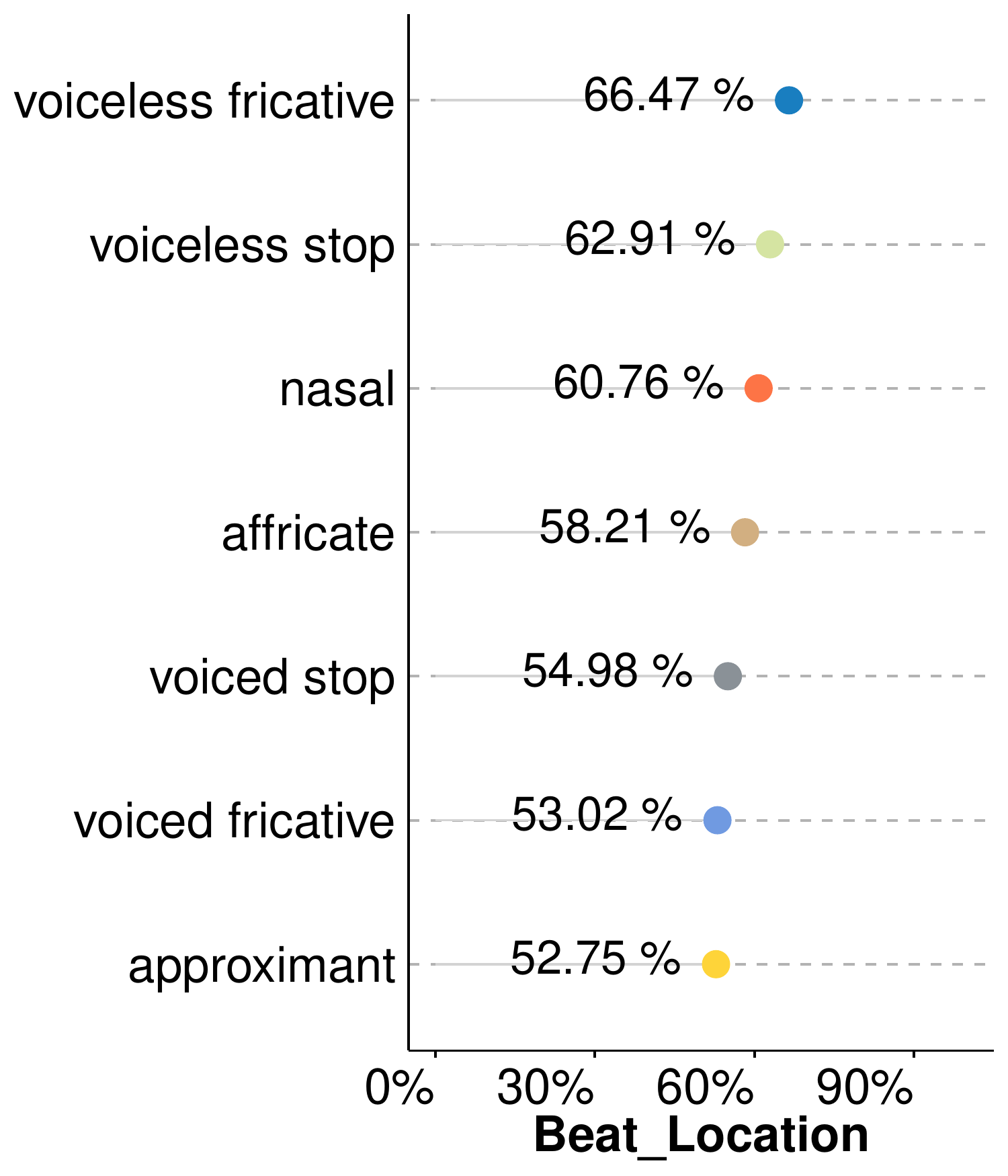}
    \includegraphics[width=0.48\linewidth]{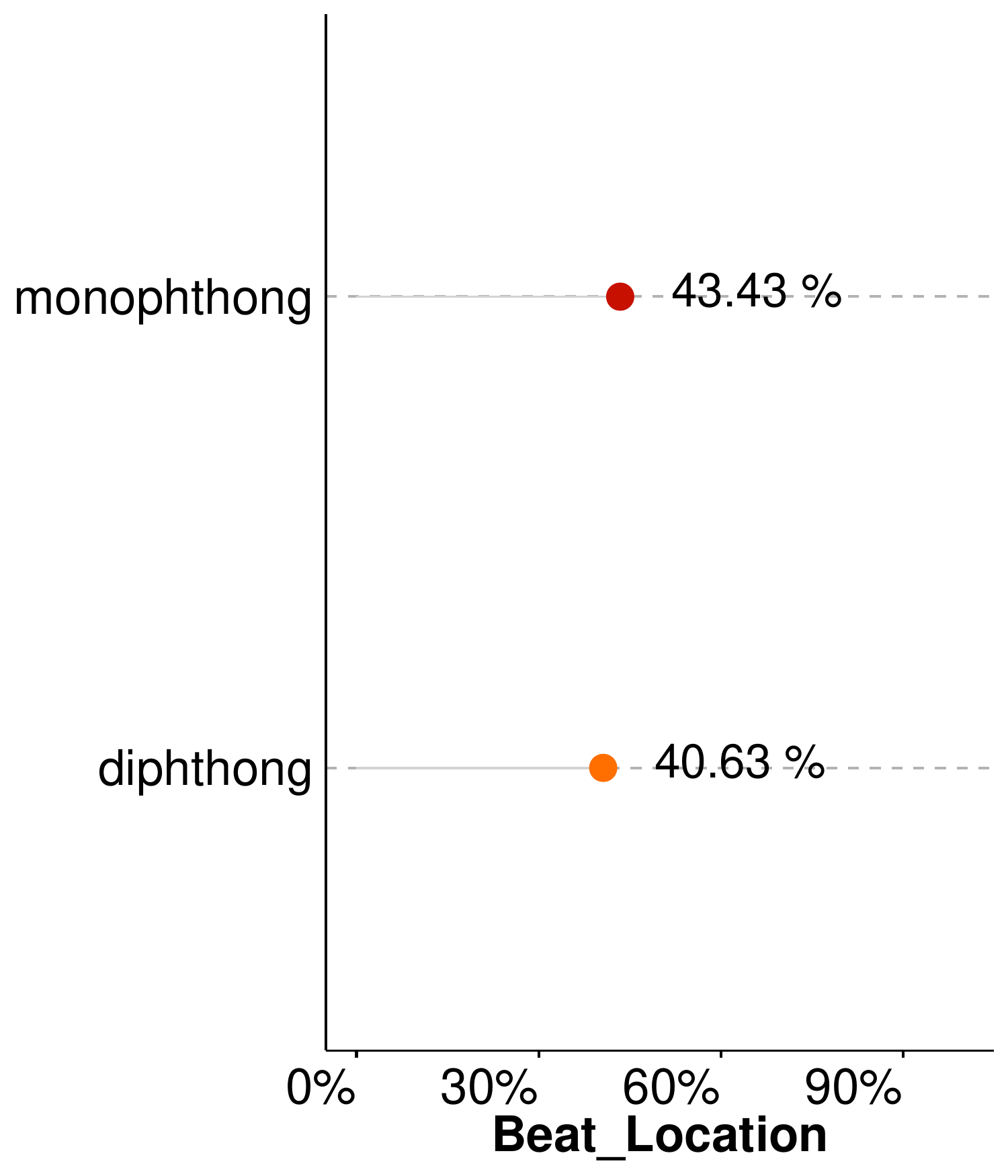}
    \caption{The average location of beat in relation to the aligned onsets ({\bf left}) and rimes ({\bf right}).}
    \label{fig:en_lolli}
\end{figure}

\textbf{Presence of Beats.} 
The English dataset consisted of 3642 English words. 2148 (59.0\%) words received musical beats. There were 10336 segments in this dataset: only 1310 segments were consonants, and the rest of 9026 were vowels. Despite the imbalance between the total numbers of consonants and vowels, the numbers of on-beat consonants and vowels were similar: 1167 consonants and 1496 vowels were aligned with beats. Figure~\ref{fig:en_bar} ranks the proportion of beat-aligned consonants and vowels from low to high. Except for voiced fricatives, all other consonant types had very high percentages of segments aligned with beats -- more than 88\%, although some had smaller absolute numbers. The sonorant phoneme types (nasals and approximants) showed high numbers of beat alignment; however, their percentages were relatively low. The other types did not seem to follow any sonority sequence either. The vowel results showed a higher percentage of diphthongs aligning with beats, which was likely to be related to phoneme length. Based on the results of beat presence from both Mandarin and English data, we can reject the hypothesis that sonorant segments are more likely to receive beats. In contrast with the Mandarin data, which had a much smaller percentage of onsets aligning with beats relative to the rimes, the English results were more unexpected. These results showed a cross-linguistic difference between English and Mandarin data in terms of whether consonants/onsets received a substantial number of beats: around 90\% of the consonants were aligned with beats in English, while only approximately 27\% of the consonants in Mandarin received beats.  \\[0.15cm]
\noindent\textbf{Location of Beats.}
Figure~\ref{fig:en_lolli} displays the beat locations for different phoneme types. The consonants show a strong correlation with the sonority hierarchy, with the exception of nasals having relatively late alignment. This may be related to the fact that a larger percentage of nasals were at syllable coda position compared with other phoneme types. The vowel data on the right did not differ much by the vowel types and were aligned at approximately the centre of the vowels. Although sonority did not substantially influence beat presence, it showed an effect on the beat location in both Mandarin and English.

%\subsubsection{xxxx}

%\subsection{English Results}

\section{Predicting Beats}
While we do not aim to provide a predictive model for use in singing synthesis, we fitted several statistical models to predict the presence of beats and the location of beats in order to test the generalisability of our findings. For each task, instead of fitting a single model, we ran the predictive models multiple times with leave-one-out cross-validation. In other words, we reserved data from a song for testing and trained the model with the rest of the data. This process was repeated multiple times until all songs were tested individually, and then the averaged performance metrics were reported below in Table~\ref{tab:beats} and~\ref{tab:location}. Unless otherwise stated, all statistical analyses were perform using the Python package \texttt{sklearn} \cite{pedregosa2011scikit}.\\[0.15cm]
\noindent\textbf{Predicting the Presence of Beats.}
Predicting the presence of beats was formulated as a binary classification task using logistic regression with default hyperparameters. Given phonetic features including duration and phoneme types, logistic regression models were trained to predict whether a beat falls on the current phone. The baseline models were simply the majority vote of the majority class (\texttt{no beat}). For each language, we fitted three models, one with duration only, one with phone types only and one with both phonetic features included. We did experiment with other features such as syllable structure and word stress but did not find them useful. Therefore they were not reported here. The averaged accuracy for each model over all test sets was presented in Table~\ref{tab:beats} below. 
In general, most models outperformed the majority vote baseline, suggesting that there are at least some regularities in terms of the placement of beats. For English, both duration and phoneme type contributed to the prediction, yet phoneme type turned out to be a stronger predictor of beat presence. In contrast, duration was the strongest predictor in the Mandarin model, whereas phoneme type was not more predictive of beat presence than the simple majority vote.\\[0.15cm]
\begin{table}[tbh]
\centering
\footnotesize
\begin{tabular}{llr}
\hline
\textbf{Language}                   & \textbf{Model}       & \textbf{Accuracy} \\ \hline
\multicolumn{1}{c}{\multirow{4}{*}{English}} & Majority vote & 0.73     \\
\multicolumn{1}{c}{}                         & Duration        & 0.79     \\
\multicolumn{1}{c}{}                         & Phone type      & 0.89     \\
\multicolumn{1}{c}{}                         & Full            & \textbf{0.92}     \\\hline
\multicolumn{1}{c}{\multirow{4}{*}{Chinese}}                     & Majority vote & 0.65     \\
                                             & Duration        & 0.76     \\
                                             & Phone type      & 0.66     \\
                                             & Full            & \textbf{0.76}    \\\hline
\end{tabular}
\caption{Predicting beat presence with logistic regression models}
\label{tab:beats}
\end{table}
\noindent\textbf{Predicting the Location of Beats.}
To predict the location of beats, we framed this as a regression task, in which regression models were employed to predict the exact location of beats within a phoneme as a percentage. For the baseline, the mean value of beat locations in the training set was used as the prediction for all beat locations in the test set. For each language, we fitted two linear regression models using duration and phoneme types respectively, and a full regression with both features. As the relations between beat locations and duration could be non-linear, we also trained regression trees on all features with a maximum tree depth of five. Averaged mean square errors (MSE) of each model are reported in Table~\ref{tab:location}.
The location of beats was more correlated to phoneme types in both languages, as the duration-only models were not outperforming the mean baseline. Yet in general, even the best performing models in both languages were not highly predictive. The full regression model in Mandarin only brought about 20\% decrease of MSE relative to the baseline, whereas the full regression tree model resulted in a 12\% decrease compared with the baseline.
\begin{table}[tbh]
\centering
\footnotesize
\begin{tabular}{clr}\hline
\multicolumn{1}{l}{\textbf{Language}} & \textbf{Model} & \textbf{Accuracy} \\\hline
\multirow{5}{*}{English}     & Mean baseline         & 0.098    \\
                             & Regression:Duration   & 0.111    \\
                             & Regression:Phone type & 0.089    \\
                             & Regression:Full       & 0.088    \\
                             & Regression Tree: Full & \textbf{0.086}    \\\hline
\multirow{5}{*}{Chinese}     & Mean baseline         & 0.105     \\
                             & Regression:Duration   & 0.129     \\
                             & Regression:Phone type & 0.086    \\
                             & Regression: Full      & \textbf{0.082}    \\
                             & Regression Tree: Full & 0.085  \\\hline 
\end{tabular}
\caption{Predicting beat location with regression models and regression trees}
\label{tab:location}
\end{table}

\section{Conclusions and Future Work}
Motivated by the genuine need to resolve bad cases in singing synthesis, this study analysed the synchronisation between speech segments and musical beats in Mandarin and English singing data. Our findings suggest that the P-centre theory of speech rhythm as well as the sonority hierarchy theory can inform segment-beat synchronisation substantially.
In summary, the research questions and hypotheses outlined in Section 1.2 can be answered as follows: (1) The sonority of the speech segments did not significantly affect the beat synchronisation. Other factors such as segment duration or syllable structure may have overridden the effect of sonority. (2) The sonority hierarchy and the P-centre theory together predicted the location of the beats: the more sonorant the segments were, the more vowel-like they were, and the earlier the alignments were. (3) Cross-linguistically, the segment-beat alignment in English and Mandarin shared some similarities (universal sonorant phoneme types), but also presented differences (much higher percentage of consonant-beat alignment in English than in Mandarin).
In the predictive models, we found that (1) the presence of beats can be predicted by segment duration (English, Mandarin) and phoneme type (English only); (2) the location of beats correlated with phoneme types in both Mandarin and English. These results exhibited cross-linguistic differences, while corroborates with the analysis in the previous sections. Further investigations can be made based on the acoustic correlates of P-centres and sonority hierarchy in order to more accurately predict speech-beat alignment, and contribute to the speech rhythm research.

\section{Acknowledgements}
Parts of this work was supported by the Summer Vacation Research Prize 2020 awarded to Cong Zhang by University of Kent. We thank Charlotte Slocombe, Matthew Cooke, and Azad Maudaressi for their assistance with data annotation.

\bibliographystyle{IEEEtran}

\bibliography{mybib}

\end{document}